\documentclass[conference]{IEEEtran}
\IEEEoverridecommandlockouts

\usepackage{amsmath,amssymb,amsfonts}

\usepackage{algorithm}
\usepackage{algpseudocode}
\usepackage{graphicx}
\usepackage[caption=false,font=footnotesize]{subfig}
\usepackage{booktabs}
\usepackage{tabularx}
\usepackage{makecell}
\usepackage{braket}
\usepackage{cite}
\usepackage{xcolor}
\usepackage{multirow}

\usepackage{orcidlink}
\usepackage{tikz}
\usepackage{pgfplots}
\pgfplotsset{compat=1.18}
\usepackage{graphicx}
\usetikzlibrary{arrows.meta, positioning, calc, fit, backgrounds}

\definecolor{DeepBlue}{HTML}{0000CD}
\definecolor{NvidiaGreen}{HTML}{76B900}
\definecolor{GoogleOrange}{HTML}{F4B400}

\pgfdeclarelayer{background}
\pgfsetlayers{background,main}

\algrenewcommand\algorithmicrequire{\textbf{Input:}}
\algrenewcommand\algorithmicensure{\textbf{Output:}}

\usepackage[top=0.74in, bottom=1in, left=0.75in, right=0.75in]{geometry}

\hypersetup{
    colorlinks=true, 
    linkcolor=black,  
    citecolor=black, 
    filecolor=black, 
    urlcolor=black    
}

\def\BibTeX{{\rm B\kern-.05em{\sc i\kern-.025em b}\kern-.08em
    T\kern-.1667em\lower.7ex\hbox{E}\kern-.125emX}}
\begin{document}

\title{Architecture-Aware Reinforcement Learning for Communication-Efficient Distributed Quantum Circuit Compilation}

\author{
\IEEEauthorblockN{
    Chien-Tung Kuo \IEEEauthorrefmark{2}\orcidlink{0009-0002-2788-9084},
    Felix Burt \IEEEauthorrefmark{3}\IEEEauthorrefmark{4}\orcidlink{0000-0002-6575-7034},
    Samuel Yen-Chi Chen \IEEEauthorrefmark{5}\orcidlink{0000-0003-0114-4826},
    Kin K. Leung\IEEEauthorrefmark{3}\orcidlink{0000-0002-3860-6257},
    Kuan-Cheng Chen\IEEEauthorrefmark{3}\IEEEauthorrefmark{4}\IEEEauthorrefmark{1}\orcidlink{0000-0002-6575-7034}}
\IEEEauthorblockA{\IEEEauthorrefmark{2}Department of Electrical Engineering, Stanford University, California, USA}
\IEEEauthorblockA{\IEEEauthorrefmark{3}Department of Electrical and Electronic Engineering, Imperial College London, London, UK}
\IEEEauthorblockA{\IEEEauthorrefmark{4}Centre for Quantum Engineering, Science and Technology (QuEST), Imperial College London, London, UK}
\IEEEauthorblockA{\IEEEauthorrefmark{5}Brookhaven National Laboratory, Upton, NY, USA}
\IEEEauthorblockA{\IEEEauthorrefmark{1} Email: kuan-cheng.chen17@imperial.ac.uk}
}

\maketitle

\begin{abstract}
Distributed quantum computing provides a scalable route for executing quantum circuits beyond the capacity limits of a single quantum processing unit (QPU), but it introduces a communication-aware compilation problem involving strict hardware constraints and circuit dependencies. This paper presents an architecture-aware reinforcement-learning framework that formulates distributed quantum compilation as a constrained Markov Decision Process (MDP). The compiler-level communication actions dynamically update logical-qubit placement and enable subsequent gate execution. A heterogeneous graph model represents interactions among hardware, logical qubits, and circuit operations, while a policy trained via Proximal Policy Optimization optimizes EPR-pair consumption and communication makespan. Evaluation across benchmark circuits shows that our policy matches state-of-the-art heuristics on structured workloads, with lookahead reward shaping yielding modest improvements on unstructured circuits.  These results demonstrate that reinforcement learning is a flexible alternative to manual heuristics, though scalability remains a key bottleneck for practical use.
\end{abstract}


\section{Introduction}
\label{sec:intro}
Scaling quantum computation beyond the resource limits of a single quantum processing unit (QPU) is a central challenge for future quantum systems. Monolithic quantum processors are limited by qubit count, physical connectivity, calibration complexity, control overhead, and accumulated noise, motivating distributed quantum computing architectures in which multiple QPUs cooperate to execute larger quantum circuits \cite{cirac1999distributed,van2016path,caleffi2024distributed}. In such systems, inter-QPU operations are enabled by entanglement-assisted communication primitives, including quantum teleportation, entanglement swapping, and non-local gate protocols based on shared entangled states \cite{bennett1993teleporting,zukowski1993event,eisert2000optimal}. While this modular approach increases effective computational scale, it introduces a new compiler-level challenge: quantum operations must be scheduled not only with respect to local hardware connectivity, but also with respect to expensive inter-QPU communication resources \cite{cacciapuoti2019quantum}.

Distributed quantum circuit compilation is therefore more complex than conventional single-device compilation. In a monolithic device, the compiler primarily addresses gate synthesis, physical-qubit mapping, and routing under local connectivity constraints \cite{shende2005synthesis,siraichi2019qubit,zulehner2018efficient,childs2019circuit,qsyn}. In a distributed architecture, the compiler must additionally determine how logical qubits are placed across QPUs, when inter-QPU communication should be introduced, and how limited communication channels should be scheduled across the network \cite{andres2019automated,sundaram2022distribution,ferrari2023modular}. Since entanglement generation and consumption are typically more costly than local quantum gates in both latency and reliability, excessive communication can increase execution time and amplify the impact of imperfect entanglement generation and noise \cite{cuomo2023optimized}. Consequently, communication volume (often measured by EPR-pair consumption) and communication latency (often measured by communication makespan) are key optimization objectives for distributed quantum compilation \cite{cuomo2023optimized,sundaram2022distribution,ferrari2023modular}.

Existing work on multi-core and distributed quantum compilation has primarily addressed qubit assignment, circuit partitioning, and teleportation-aware routing \cite{andres2019automated,wu2022autocomm,wu2023qucomm,sundaram2022distribution,chen2026adaptive,mao2023qubit}. Recent heuristic mapping algorithms optimize initial placements using assignment formulations \cite{escofet2025revisiting} or adapt SABRE to jointly incorporate intra-core SWAPs, qubit teleportation, and gate teleportation \cite{russo2025telesabre}. Complementary partitioning frameworks combine circuit-structure analysis with communication-cost scheduling to optimize logical qubit placement and non-local operation realization \cite{burt2024generalised,burt2026multilevel,andres2024distributing,burt2025entanglement}. Together, these structural and mapping approaches establish that initial placement and primitive selection substantially impact communication overhead.

Parallel efforts have focused on analytical characterization, network scheduling, and learning-based compilation. Graph-theoretic circuit profiling \cite{bandic2025profiling} and inter-core traffic characterizations \cite{benrached2025traffic} provide insights into communication bottlenecks and hardware-algorithm matching. Meanwhile, network-aware scheduling accounts for probabilistic entanglement generation, contention, and capacity constraints during execution \cite{pouryousef2025network}. Recent reinforcement learning (RL) formulations target adaptive compilation under stochastic EPR dynamics \cite{promponas2025compiler}. While these advances address isolated mapping or scheduling constraints, jointly optimizing communication decisions across dynamic topologies, circuit dependencies, and hardware capacity remains an open challenge.

To address this challenge, we propose a heterogeneous graph reinforcement learning framework for communication-aware distributed quantum compilation. Our policy sequentially selects compiler-level communication actions under dynamic placement and capacity constraints. By representing compiler states as heterogeneous graphs, the policy reasons over circuit dependencies and hardware availability to optimize long-horizon communication decisions.

The main contributions of this work are:
\begin{itemize}
    \item We propose a RL framework that models communication scheduling and dynamic qubit placement as a constrained Markov Decision Process (MDP).
    \item We introduce explicit dynamic constraints on active EPR operations, enforcing physical capacity limits on concurrent communication primitives.
    \item We construct a heterogeneous graph state representation that encodes circuit dependencies, qubit placement, and QPU topology.
    \item We demonstrate competitive compilation performance against state-of-the-art heuristics across diverse benchmark circuits.
\end{itemize}

\section{Background and Preliminaries}\label{sec:preliminaries}
This section introduces the circuit, architecture, and communication abstractions used throughout the paper. We first introduce the entanglement-assisted split and merge primitives used for non-local operations, and proceed to describe the distributed quantum circuit model and logical-qubit availability across QPUs. We then formulate distributed compilation as a MDP.

\subsection{Entanglement-Assisted Communication}

Inter-QPU operations are enabled by entanglement-assisted communication primitives, such as teleportation, entanglement swapping, and non-local gate protocols based on shared entangled states \cite{bennett1993teleporting,eisert2000optimal,yimsiriwattana2004generalized}. In this work, such mechanisms are abstracted at the compiler level as communication actions that modify logical-qubit availability across QPUs. This abstraction is used only for scheduling and placement decisions; it does not imply physical cloning of an unknown quantum state~\cite{wootters1982single}.

\begin{figure}[t]
    \centering
    \includegraphics[]{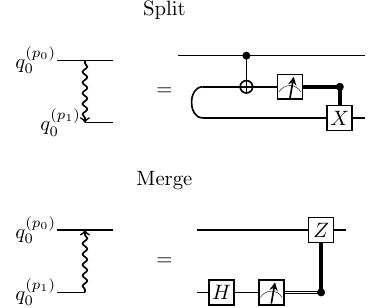}
    \caption{The two communication primitives. A split operation uses a Bell state (EPR pair) to make a qubit non-locally accessible as a control. A merge removes one of the two qubits from the joint state. Teleportation circuits can be built by composing splits and merges.}
    \label{fig:communication}
\end{figure}

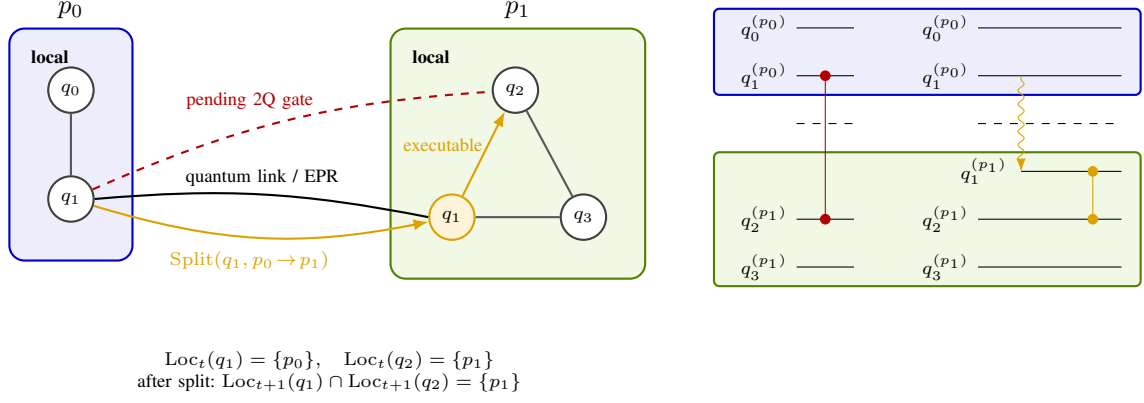
\begin{figure*}[t]
\centering
\begin{tikzpicture}[
    scale=1.2,
    font=\scriptsize,
    >=Latex,
    qubit/.style={
        circle,
        draw=black!75,
        fill=white,
        minimum size=6mm,
        inner sep=0pt,
        thick
    },
    qpuboxblue/.style={
        draw=DeepBlue,
        fill=DeepBlue!7,
        rounded corners=2mm,
        thick,
        inner sep=5mm
    },
    qpuboxgreen/.style={
        draw=NvidiaGreen!70!black,
        fill=NvidiaGreen!10,
        rounded corners=2mm,
        thick,
        inner sep=5mm
    },
    local/.style={thick, draw=black!65},
    remote/.style={dashed, thick, draw=red!70!black},
    quantumlink/.style={
        thick,
        draw=black
    },
    comm/.style={-{Latex[length=2mm]}, thick, draw=GoogleOrange!90!black}
]

\node[qubit] (q0) at (0,0.95) {$q_0$};
\node[qubit] (q1) at (0,-0.25) {$q_1$};

\node[qubit] (q2) at (4.90,0.95) {$q_2$};
\node[qubit, draw=GoogleOrange!90!black, fill=GoogleOrange!15] (q1copy) at (4.20,-0.45) {$q_1$};
\node[qubit] (q3) at (5.65,-0.45) {$q_3$};

\begin{scope}[on background layer]
\node[qpuboxblue, fit=(q0)(q1), label={[font=\bfseries]above:$p_0$}] (p0) {};
\node[qpuboxgreen, fit=(q2)(q3)(q1copy), label={[font=\bfseries]above:$p_1$}] (p1) {};

\end{scope}

\node[anchor=north west, font=\scriptsize\bfseries]
    at ($(p0.north west)+(1.5mm,-1.5mm)$) {local};

\node[anchor=north west, font=\scriptsize\bfseries]
    at ($(p1.north west)+(1.5mm,-1.5mm)$) {local};

\draw[local] (q0) -- (q1);
\draw[local] (q2) -- (q3);
\draw[local] (q3) -- (q1copy);

\draw[remote]
    (q1) to[bend left=10]
    node[above, pos=0.46, xshift=-3mm, yshift=1.5mm,
         text=red!70!black, fill=white, inner sep=1pt]
    {\scriptsize pending 2Q gate}
    (q2);

\draw[quantumlink]
    (q1.east) to[bend left=8]
    node[above, pos=0.50, yshift=0.6mm,
         text=black, inner sep=1pt]
    {\scriptsize quantum link / EPR}
    (q1copy.west);

\draw[comm]
    (q1) to[bend right=14]
    node[below, pos=0.43, xshift=1.5mm, yshift=-1.5mm,
         text=GoogleOrange!90!black, fill=white, inner sep=1pt]
    {\scriptsize $\mathrm{Split}(q_1,p_0\!\rightarrow\!p_1)$}
    (q1copy);

\draw[comm]
    (q1copy) --
    node[pos=0.52, xshift=-5.5mm, yshift=1mm,
         text=GoogleOrange!90!black, inner sep=1pt]
    {\scriptsize executable}
    (q2);

\node[align=center, font=\scriptsize] at (2.85,-2.15)
{$\mathrm{Loc}_{t}(q_1)=\{p_0\},\quad \mathrm{Loc}_{t}(q_2)=\{p_1\}$\\
after split: $\mathrm{Loc}_{t+1}(q_1)\cap \mathrm{Loc}_{t+1}(q_2)=\{p_1\}$};

\begin{scope}[xshift=8cm,yshift=-1cm,scale=1.000000,x=1pt,y=1pt]
\draw[color=black] (0.000000,75.000000) -- (18.000000,75.000000);
\draw[color=black] (0.000000,75.000000) node[left] {$q_{0}^{(p_0)}$};
\draw[color=black] (0.000000,60.000000) -- (18.000000,60.000000);
\draw[color=black] (0.000000,60.000000) node[left] {$q_{1}^{(p_0)}$};
\draw[color=black,dashed] (0.000000,45.000000) -- (18.000000,45.000000);
\draw[color=black] (0.000000,15.000000) -- (18.000000,15.000000);
\draw[color=black] (0.000000,15.000000) node[left] {$q_{2}^{(p_1)}$};
\draw[color=black] (0.000000,0.000000) -- (18.000000,0.000000);
\draw[color=black] (0.000000,0.000000) node[left] {$q_{3}^{(p_1)}$};
\draw[color=red!70!black] (9.000000,60.000000) -- (9.000000,15.000000);
\filldraw[color=red!70!black] (9.000000, 60.000000) circle(1.500000pt);
\begin{scope}
\filldraw[color=red!70!black]  (9.000000, 15.000000) circle(1.500000pt);
\clip (9.000000, 15.000000) circle(3.000000pt);
\end{scope}
\end{scope}

\begin{scope}[xshift=10cm,yshift=-1cm,scale=1.000000,x=1pt,y=1pt]
\draw[color=black] (0.000000,75.000000) -- (45.000000,75.000000);
\draw[color=black] (0.000000,75.000000) node[left] {$q_{0}^{(p_0)}$};
\draw[color=black] (0.000000,60.000000) -- (45.000000,60.000000);
\draw[color=black] (0.000000,60.000000) node[left] {$q_{1}^{(p_0)}$};
\draw[color=black,dashed] (0.000000,45.000000) -- (45.000000,45.000000);
\draw[color=black] (13.500000,30.000000) -- (45.000000,30.000000);
\draw[color=black] (0.000000,15.000000) -- (45.000000,15.000000);
\draw[color=black] (0.000000,15.000000) node[left] {$q_{2}^{(p_1)}$};
\draw[color=black] (0.000000,0.000000) -- (45.000000,0.000000);
\draw[color=black] (0.000000,0.000000) node[left] {$q_{3}^{(p_1)}$};
\draw[decorate,color=GoogleOrange!90!black,decoration={snake,amplitude=.4mm,segment length=1.8mm, post length=0.5mm, pre length=0.5mm},->] (13.500000,60.000000) -- (13.500000,30.000000);
\draw[color=black] (12.000000,30.000000) node[left] {$q_{1}^{(p_1)}$};
\draw[color=GoogleOrange!90!black] (36.000000,30.000000) -- (36.000000,15.000000);
\filldraw[color=GoogleOrange!90!black] (36.000000, 30.000000) circle(1.500000pt);
\begin{scope}
\filldraw[color=GoogleOrange!90!black] (36.000000, 15.000000) circle(1.500000pt);
\end{scope}
\end{scope}

\begin{scope}[on background layer]
\draw[draw=DeepBlue, fill=DeepBlue!7, rounded corners=2pt, thick]
    (7.3cm-6pt,-1cm+54pt) rectangle (10cm+51pt,-1cm+81pt);
\draw[draw=NvidiaGreen!70!black, fill=NvidiaGreen!10, rounded corners=2pt, thick]
    (7.3cm-6pt,-1cm-6pt) rectangle (10cm+51pt,-1cm+36pt);
\end{scope}

\end{tikzpicture}
\caption{Example of a distributed two-qubit gate. Initially, $q_1$ and $q_2$
are available on different QPUs, so the gate cannot be executed locally.
An inter-QPU quantum link provides the entanglement-assisted communication
resource used by the compiler-level split action. The split makes $q_1$
available on $p_1$ at the compiler level, enabling local execution with
$q_2$, without implying physical cloning of an unknown quantum state. After the gate has been executed, $q_1$ can either remain split for future gates,  or merge to either $p_0$ or $p_1$.}
\label{fig:distributed_circuit_example}
\end{figure*}

We use two compiler-level primitives shown in Fig.~\ref{fig:communication}. A \emph{split} action consumes one EPR pair to make a logical qubit available at an additional QPU for distributed controlled-unitary operations. A \emph{merge} action removes redundant distributed placement when it is no longer useful. The split and merge operations are also known as the \textit{cat-entangler} and \textit{disentangler} operations~\cite{yimsiriwattana2004generalized} or the entanglement-assisted \textit{starting} and \textit{ending processes}~\cite{wu2023entanglement}. As illustrated in Fig.~\ref{fig:distributed_circuit_example}, a split can enable an otherwise non-local two-qubit gate, while the subsequent merge decision affects the placement and communication opportunities of later gates. Using a symmetric two-qubit $CP$ gate  means that non-local gates can be executed by splitting either control, target, or both.

\subsection{Distributed Quantum Circuit Model}
A quantum circuit is represented as a dependency-constrained collection of quantum operations acting on a set of logical qubits. Let
\begin{equation}
    \mathcal{C}=(Q,\mathcal{G},\prec)
\end{equation}
denote a circuit, where $Q$ is the set of logical qubits, $\mathcal{G}$ is the set of gates, and $\prec$ denotes the partial order induced by gate dependencies. A gate may be executed only after all of its predecessors have been completed. Single-qubit gates act on one logical qubit, whereas two-qubit gates require both operands to be simultaneously available at a valid execution location. We do not place requirements on single-qubit gates, though require that two-qubit $CP$ gates be symmetric.

In distributed quantum computing, the circuit is executed over multiple quantum processing units (QPUs). We model the target architecture as a communication graph
\begin{equation}
    \mathcal{H}=(P,E),
\end{equation}
where $P$ is the set of QPUs and $E$ is the set of inter-QPU communication links. Each QPU $p\in P$ has a finite qubit capacity and a finite number of communication channels. At compilation step $t$, the placement of a logical qubit $q$ is described by
\begin{equation}
    \mathrm{Loc}_{t}(q)\subseteq P,
\end{equation}
which denotes the set of QPUs at which the compiler currently regards $q$ as available for distributed execution. Note that the location is a subset of all partitions since a qubit can in principle be split across all QPUs. A two-qubit gate acting on $q_i$ and $q_j$ can be executed locally when there exists a QPU $p$ such that $p\in \mathrm{Loc}_{t}(q_i)\cap \mathrm{Loc}_{t}(q_j)$. Otherwise, inter-QPU communication is required before the gate can be executed \cite{siraichi2019qubit,sundaram2022distribution}.

\begin{figure}[!b]
\centering
\begin{tikzpicture}[
    font=\scriptsize,
    >=Latex,
    node distance=8mm,
    box/.style={
        draw,
        rounded corners=1.5mm,
        align=center,
        minimum width=2.7cm,
        minimum height=0.85cm,
        inner sep=2mm
    },
    flow/.style={-{Latex[length=2mm]}, thick}
]

\definecolor{DeepBlue}{HTML}{0000CD}
\definecolor{SoftBlue}{HTML}{E8EAFE}
\definecolor{NvidiaGreen}{HTML}{76B900}
\definecolor{SoftGreen}{HTML}{EEF7E5}
\definecolor{GoogleOrange}{HTML}{F4B400}
\definecolor{SoftOrange}{HTML}{FFF4D6}
\definecolor{SoftGray}{HTML}{F8F8F8}

\node[box, draw=DeepBlue, fill=SoftBlue] (circuit)
{\textbf{Circuit DAG}\\ qubits, gates, dependencies};

\node[box, draw=NvidiaGreen!85!black, fill=SoftGreen, right=9mm of circuit] (hardware)
{\textbf{QPU Network}\\ topology, capacity, channels};

\node[box, draw=GoogleOrange!90!black, fill=SoftOrange, below=8mm of $(circuit)!0.5!(hardware)$] (comm)
{\textbf{Communication Action}\\ split / merge};

\node[box, draw=black!70, fill=SoftGray, below=8mm of comm] (state)
{\textbf{Compiler State}\\ placement, clocks, frontier};

\draw[flow] (circuit.south) |- (comm.west);
\draw[flow] (hardware.south) |- (comm.east);
\draw[flow] (comm.south) -- (state.north);

\coordinate (stateleft) at ($(state.west)+(-15mm,0)$);
\coordinate (circuitentry) at ($(circuit.south)+(-10mm,0)$);
\coordinate (feedbacktop) at ($(stateleft |- circuitentry)+(0,-8mm)$);

\draw[flow, dashed]
    (state.west) -- (stateleft)
    -- (feedbacktop)
    -- (circuitentry);

\end{tikzpicture}%
\caption{Preliminary abstraction of distributed quantum circuit compilation. A circuit DAG and distributed-QPU architecture define a compiler state; communication actions update placement and resource availability.}
\label{fig:prelim_abstraction}
\end{figure}
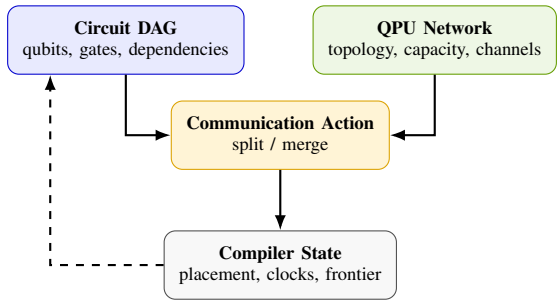

\subsection{Sequential Compilation Formulation}

Distributed quantum circuit compilation is naturally sequential because a communication action may affect both immediate gate execution and future communication opportunities. We therefore formulate the compilation process as a Markov decision process (MDP)
\begin{equation}
(\mathcal{S},\mathcal{A},\mathcal{T},r),
\end{equation}
where $\mathcal{S}$, $\mathcal{A}$, $\mathcal{T}$, and $r$ denote the state space, communication-action space, transition function, and reward function, respectively. At each step, the compiler selects a split or merge action based on the current circuit, qubit-placement, and hardware-resource state.

\section{Problem Formulation and System Model}
\label{sec:problem}
Building upon the distributed circuit abstraction introduced in Section~\ref{sec:preliminaries}, we formulate distributed quantum compilation as an optimization problem over communication scheduling and dynamic logical-qubit placement. The compiler must determine when communication actions should occur and how logical-qubit availability should evolve across QPUs so that all circuit dependencies are satisfied under hardware constraints.

\subsection{Problem Definition}

Given a quantum circuit
\(
\mathcal{C}=(Q,\mathcal{G},\prec)
\)
and a distributed QPU architecture
\(
\mathcal{H}=(P,E),
\)
the compiler seeks a sequence of communication actions
\begin{equation}
\Pi=(a_1,a_2,\ldots,a_K)
\end{equation}
that enables the execution of all gates in $\mathcal{G}$.

Each action is a feasible split or merge operation that updates the compiler-level availability of logical qubits across QPUs:
\begin{equation}
\mathrm{Loc}_{t}(q)
\rightarrow
\mathrm{Loc}_{t+1}(q).
\end{equation}
The resulting compilation specifies an evolving logical-qubit placement and an inter-QPU communication schedule while preserving circuit dependencies and hardware constraints.

\subsection{Optimization Objective}

\label{sec:opt_obj}

Among feasible compilations, the goal of distributed quantum compilation is to reduce the communication overhead introduced by inter-QPU execution. We primarily evaluate communication volume using the number of EPR pairs consumed, which is equivalent to the number of split operations under our communication model. 

Communication makespan is also important because longer execution times increase exposure to decoherence. However, since makespan depends strongly on the assumed communication and entanglement-generation models, we use it as an internal reward signal to guide compilation rather than as a metric for cross-method comparison.

\subsection{System Model}

Here we specify the resource and temporal constraints used during compilation.

\subsubsection{Resource Constraints}

Each QPU $p \in P$ has finite computational and communication resources. Its qubit capacity limits the number of logical qubits that may be simultaneously available at $p$, while its communication-channel capacity limits the number of split actions that can be scheduled from $p$ within one communication round. These constraints restrict both communication parallelism and placement flexibility.

Formally, the qubit-capacity constraint requires
\begin{equation}
    |\{q \in Q : p \in \mathrm{Loc}_{t}(q)\}|
    \leq
    \mathrm{capacity}[p],
\end{equation}
for every QPU $p \in P$.

The communication-channel constraint requires that the number of split actions scheduled from QPU $p$ within one communication round does not exceed
\begin{equation}
    \mathrm{max\_channels}[p].
\end{equation}

\subsubsection{Temporal Communication Model}

\label{sec:temporal_comm}

We adopt a compiler-level round-based communication model. Each QPU $p\in P$ maintains a logical communication clock
\(
\mathrm{clock}[p]
\)
and a remaining channel budget
\(
\mathrm{active\_channels}[p].
\)
The clock records the number of communication rounds required by QPU $p$, while the channel counter records how many split actions from $p$ can still be scheduled in the current round.

For a split action
\(
\mathrm{Split}(q,A\rightarrow B),
\)
the compiler schedules an entanglement-assisted communication event that makes logical qubit $q$ available for distributed execution involving QPU $B$. This action consumes one communication channel from the source QPU $A$.

If $A$ has at least one available channel, the split is scheduled in the current communication round:
\begin{equation}
    \mathrm{active\_channels}[A]
    \leftarrow
    \mathrm{active\_channels}[A]-1.
\end{equation}

If no channel is available at $A$, the communication clock of $A$ is advanced by one round, its channel budget is refreshed, and one channel is consumed:
\begin{equation}
    \mathrm{clock}[A]
    \leftarrow
    \mathrm{clock}[A]+1,
\end{equation}
\begin{equation}
    \mathrm{active\_channels}[A]
    \leftarrow
    \mathrm{max\_channels}[A]-1.
\end{equation}

This model captures contention for limited communication bandwidth while keeping the transition model lightweight for reinforcement learning. The communication makespan is defined as
\begin{equation}
    T_{\mathrm{comm}}
    =
    \max_{p\in P}\mathrm{clock}[p].
\end{equation}

\subsection{Feasibility Constraints}

A valid distributed compilation must satisfy gate-execution, resource, topology, and dependency constraints.

\subsubsection{Communication-Action Feasibility}

Communication actions must satisfy topology, channel, and QPU capacity constraints. A split action
\(
\mathrm{Split}(q,A\rightarrow B)
\)
is feasible only if the QPUs are connected in the communication graph
\begin{equation}
    (A,B)\in E,
\end{equation}
and both QPUs have available hardware communication channels to initiate the link
\begin{equation}
    \mathrm{avail\_channels}[A] > 0 
    \quad \text{and} \quad 
    \mathrm{avail\_channels}[B] > 0.
\end{equation}
Unlike standard mapping frameworks, explicitly enforcing channel capacity models dynamic link contention during batched EPR pair generation. Splitting does not consume physical qubit slots at $B$, and updates placement as 
\begin{equation}
    \mathrm{Loc}_{t+1}(q) = \mathrm{Loc}_{t}(q)\cup\{B\}.
    \label{eq:split}
\end{equation}

A merge action $\mathrm{Merge}(q,A)$ resolves all EPR pairs involving qubit $q$ at QPU $A$, updating placement to 
\begin{equation}
    \mathrm{Loc}_{t+1}(q) = \mathrm{Loc}_{t}(q)\setminus\{A\}
    \label{eq:merge}
\end{equation} and releasing the associated channels back into the pool. If merging at the source QPU $A$ teleports state to a target QPU $B$, the merge is feasible only if $B$ has sufficient physical qubit capacity:
\begin{equation}
    |\{q' \in Q : B \in \mathrm{Loc}_{t}(q')\}| < \mathrm{capacity}[B].
\end{equation}

\subsubsection{Gate Execution Feasibility}

A gate $g \in \mathcal{G}$ may execute only when all required operands are available at a valid execution location and all predecessor gates have completed. Restriction to symmetric two-qubit gates means any gate acting on logical qubits $q_i$ and $q_j$ can be executed if there exists a QPU $p \in P$ such that
\begin{equation}
    p \in \mathrm{Loc}_{t}(q_i) \cap \mathrm{Loc}_{t}(q_j).
\end{equation}

For single qubit gates, a logical qubit is require to be uniquely localized:
\begin{equation}
    |\mathrm{Loc}_{t}(q)| = 1.
\end{equation}
This is because non-diagonal single-qubit gates cannot be executed while a qubit is split~\cite{wu2023entanglement}. For simplicity, we enforce the rule for diagonal gates too.

\section{Reinforcement Learning Framework Design}
\label{sec:rl}

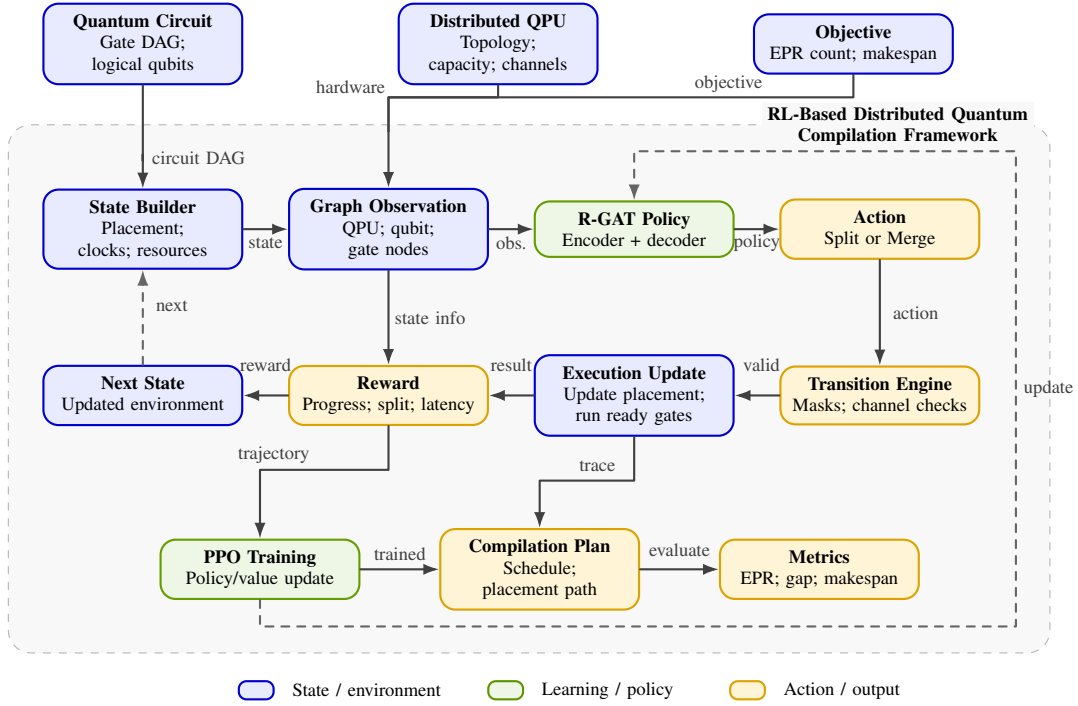
\begin{figure*}[t]
\centering
\begin{tikzpicture}[
    >=Latex,
    line join=round,
    line cap=round
]

\definecolor{DeepBlue}{HTML}{0000CD}
\definecolor{NvidiaGreen}{HTML}{76B900}
\definecolor{GoogleOrange}{HTML}{F4B400}

\definecolor{SoftBlue}{HTML}{E8EAFE}
\definecolor{SoftGreen}{HTML}{EEF7E5}
\definecolor{SoftOrange}{HTML}{FFF4D6}
\definecolor{SoftGray}{HTML}{F8F8F8}

\tikzset{
    box/.style={
        rounded corners=2mm,
        thick,
        align=center,
        inner sep=1.4mm,
        minimum height=0.62cm,
        text width=2.35cm,
        font=\scriptsize
    },
    input/.style={box, draw=DeepBlue, fill=SoftBlue},
    state/.style={box, draw=DeepBlue, fill=SoftBlue},
    learn/.style={box, draw=NvidiaGreen!85!black, fill=SoftGreen},
    action/.style={box, draw=GoogleOrange!90!black, fill=SoftOrange},
    flow/.style={-{Latex[length=2.1mm,width=1.5mm]}, thick, draw=black!75},
    route/.style={thick, draw=black!75},
    feedback/.style={-{Latex[length=2.1mm,width=1.5mm]}, thick, dashed, draw=black!60},
    group/.style={
        draw=black!30,
        dashed,
        rounded corners=3mm,
        inner xsep=4.5mm,
        inner ysep=3.5mm,
        fill=SoftGray
    },
    title/.style={
        font=\bfseries\scriptsize,
        fill=SoftGray,
        inner sep=2pt,
        align=center
    },
    lab/.style={
        font=\scriptsize,
        inner sep=0.3pt,
        text=black!75
    },
    legendbox/.style={
        rounded corners=1mm,
        minimum width=0.46cm,
        minimum height=0.24cm,
        thick
    }
}

\node[input] (circuit) at (0,0) {
\textbf{Quantum Circuit}\\
Gate DAG; logical qubits
};

\node[input] (arch) at (4.7,0) {
\textbf{Distributed QPU}\\
Topology; capacity; channels
};

\node[input] (obj) at (9.4,0) {
\textbf{Objective}\\
EPR count; makespan
};

\node[state] (statebuilder) at (0,-2.45) {
\textbf{State Builder}\\
Placement; clocks; resources
};

\node[state] (graph) at (3.25,-2.45) {
\textbf{Graph Observation}\\
QPU; qubit; gate nodes
};

\node[learn] (policy) at (6.5,-2.45) {
\textbf{R-GAT Policy}\\
Encoder + decoder
};

\node[action] (act) at (9.75,-2.45) {
\textbf{Action}\\
Split or Merge
};

\node[action] (transition) at (9.75,-4.65) {
\textbf{Transition Engine}\\
Masks; channel checks
};

\node[state] (update) at (6.5,-4.65) {
\textbf{Execution Update}\\
Update placement; run ready gates
};

\node[action] (reward) at (3.25,-4.65) {
\textbf{Reward}\\
Progress; split; latency
};

\node[state] (nextstate) at (0,-4.65) {
\textbf{Next State}\\
Updated environment
};

\node[learn] (ppo) at (1.55,-6.95) {
\textbf{PPO Training}\\
Policy/value update
};

\node[action] (plan) at (5.25,-6.95) {
\textbf{Compilation Plan}\\
Schedule; placement path
};

\node[action] (metrics) at (8.95,-6.95) {
\textbf{Metrics}\\
EPR; gap; makespan
};

\coordinate (updateRightBottom) at ($(ppo.south)+(10.00,-0.35)$);
\coordinate (updateRightTop) at ($(policy.north)+(5.05,0.65)$);
\coordinate (policyTopEntry) at ($(policy.north)+(0,0.65)$);

\begin{pgfonlayer}{background}
\node[group,
    fit=(statebuilder)(graph)(policy)(act)(transition)(update)(reward)(nextstate)
        (ppo)(plan)(metrics)(updateRightBottom)(updateRightTop)(policyTopEntry)
] (framework) {};
\end{pgfonlayer}

\node[title, anchor=north east]
    at ($(framework.north east)+(-0.22,+0.3)$)
{RL-Based Distributed Quantum\\Compilation Framework};

\coordinate (entryA) at ($(statebuilder.north)+(0,0.45)$);

\coordinate (graphMerge) at ($(graph.north)+(0,0.75)$);

\draw[flow] (circuit.south) -- ++(0,-1.00)
    -- node[lab, right, xshift=1.0mm] {circuit DAG}
    (entryA) -- (statebuilder.north);

\draw[route] (arch.south) -- ++(0,-0.16)
    -| (graphMerge);
\node[lab, above, xshift=-9.5mm,yshift=1mm]
    at ($(arch.south)+(-1,-0.22)$) {hardware};

\draw[route] (obj.south) -- ++(0,-0.30)
    -| (graphMerge);
\node[lab, above, xshift=-6.5mm]
    at ($(obj.south)+(-1,-0.25)$) {objective};

\draw[flow] (graphMerge) -- (graph.north);

\draw[flow] (graph.south) --
    node[lab, midway, right, xshift=0.6mm] {state info}
    (reward.north);

\draw[flow] (statebuilder.east) --
    node[lab, midway, yshift=-1.8mm] {state}
    (graph.west);

\draw[flow] (graph.east) --
    node[lab, midway, yshift=-1.8mm] {obs.}
    (policy.west);

\draw[flow] (policy.east) --
    node[lab, midway, yshift=-1.8mm] {policy}
    (act.west);

\draw[flow] (act.south) --
    node[lab, right, xshift=1.5mm] {action}
    (transition.north);

\draw[flow] (transition.west) --
    node[lab, above, yshift=3.0mm, xshift=0.6mm] {valid}
    (update.east);

\draw[flow] (update.west) --
    node[lab, above, yshift=3.0mm] {result}
    (reward.east);

\draw[flow] (reward.west) --
    node[lab, above, yshift=3.0mm] {reward}
    (nextstate.east);

\draw[feedback] (nextstate.north) -- ++(0,0.58)
    -| (statebuilder.south)
    node[lab, pos=0.25, right, yshift=2.0mm, xshift=+1.5mm] {next};

\draw[flow] (reward.south) -- ++(0,-0.58)
    -| (ppo.north)
    node[lab, pos=0.30, left, yshift=2.0mm] {trajectory};

\draw[feedback]
    (ppo.south) -- ++(0,-0.35)
    -| (updateRightBottom)
    -- node[lab, right, xshift=0.8mm] {update} (updateRightTop)
    -- (policyTopEntry)
    -- (policy.north);

\draw[flow] (update.south) -- ++(0,-0.58)
    -| (plan.north)
    node[lab, pos=0.30, right, yshift=2.0mm] {trace};

\draw[flow] (ppo.east) --
    node[lab, above,yshift=1.0mm] {trained}
    (plan.west);

\draw[flow] (plan.east) --
    node[lab, above,yshift=1.5mm] {evaluate}
    (metrics.west);

\node[legendbox, draw=DeepBlue, fill=SoftBlue] (leg1) at (1.5,-8.55) {};
\node[anchor=west, font=\scriptsize] at ($(leg1.east)+(0.11,0)$)
{State / environment};

\node[legendbox, draw=NvidiaGreen!80!black, fill=SoftGreen] (leg2) at (4.8,-8.55) {};
\node[anchor=west, font=\scriptsize] at ($(leg2.east)+(0.11,0)$)
{Learning / policy};

\node[legendbox, draw=GoogleOrange!90!black, fill=SoftOrange] (leg3) at (8.0,-8.55) {};
\node[anchor=west, font=\scriptsize] at ($(leg3.east)+(0.11,0)$)
{Action / output};

\end{tikzpicture}%
\caption{Overview of the proposed RL framework. The state builder constructs a graph
observation from circuit structure, QPU topology, current placement, and optional
objective-conditioning features. The R-GAT policy processes this observation
and selects a feasible split or merge action. The transition
engine applies topology, placement, capacity, channel, and dependency
constraints, after which executable gates are deterministically updated.
Rewards are computed from communication cost, circuit progress, lookahead
executability, and terminal completion. PPO updates the policy and value
networks from collected trajectories.}
\label{fig:rl_framework}
\end{figure*}

To address the distributed compilation problem introduced in Section~\ref{sec:problem}, we model communication scheduling as a MDP. At each step, the compiler selects communication actions that modify qubit placement, update resource availability, and enable future gate execution under distributed hardware constraints.

The proposed framework combines graph-based state representation, reinforcement learning, and hierarchical action selection to efficiently explore the large combinatorial space induced by distributed quantum compilation.

\subsection{MDP Formulation}
\label{sec:MDP}

In order to apply RL, we must formulate distributed quantum compilation as a MDP \cite{rl_mdp}. Formally, the MDP is defined as
\(
(\mathcal{S}, \mathcal{A}, \mathcal{T}, r)
\),
where \(\mathcal{S}\) denotes the state space, \(\mathcal{A}\) denotes the communication-action space, \(\mathcal{T}\) defines state transitions, and \(r\) specifies the reward function. Each episode begins from an initial compiler state and terminates when all circuit operations have been executed.

\subsubsection{State Space \(\mathcal{S}\)}

A state
\(
s_t \in \mathcal{S}
\)
represents the compiler state at step \(t\). Following the distributed compilation abstraction introduced in Section~II, the state includes:

\begin{itemize}
    \item current qubit placement
    \(
    \mathrm{Loc}_t(q) 
    \)
    for all logical qubits \(q \in Q\)

    \item communication clocks
    \(
    \mathrm{clock}[p]
    \)
    for each QPU \(p \in P\),

    \item remaining communication channels and qubit capacity

    \item circuit progress, including completed gates and executable frontier
\end{itemize}

Together, these variables define the information required to evaluate future communication decisions.

\subsubsection{Action Space \(\mathcal{A}\)}

The action space consists of two compiler-level communication primitives:
\begin{itemize}
    \item \(\mathrm{Split}(q,A\rightarrow B)\), which schedules an entanglement-assisted communication event from source QPU \(A\) to destination QPU \(B\), making \(q\) available at \(B\).
    \item \(\mathrm{Merge}(q,A)\), which removes redundant placement of \(q\) at QPU \(A\).
\end{itemize}
Actions are restricted by topology, placement, capacity, and channel constraints.

\subsubsection{Transition Function \(\mathcal{T}\)}

Given state \(s_t\) and action \(a_t\), the transition function
\begin{equation}
    s_{t+1} = \mathcal{T}(s_t,a_t)
\end{equation}
updates qubit placement, communication resources, QPU clocks, and circuit progress.

For a split action \(\mathrm{Split}(q,A\rightarrow B)\), the transition first applies the round-based source-channel update defined in Section~\ref{sec:temporal_comm}. The placement set is then updated according to Equations \ref{eq:split} and \ref{eq:merge}. After each communication action, all gates whose dependency and placement constraints are satisfied are automatically executed. 

\subsubsection{Reward Function \(r\)}

The reward function aligns policy learning with the communication-aware
optimization objective introduced in Section~\ref{sec:opt_obj}. A single
scalar objective provides only sparse, delayed feedback and cannot
distinguish necessary communication from avoidable communication, making
credit assignment difficult over long compilation trajectories. We
therefore decompose the reward into five complementary terms, each
targeting a distinct aspect of compilation quality: temporal contention,
communication cost, immediate progress, delayed locality benefit, and
task completion. The step reward is the weighted combination

\begin{equation}
\begin{split}
R_t = w_{\mathrm{lat}} r_{\mathrm{lat}} &+ w_{\mathrm{split}} r_{\mathrm{split}}
+ w_{\mathrm{prog}} r_{\mathrm{prog}} \\&+ w_{\mathrm{look}} r_{\mathrm{look}}
+ w_{\mathrm{comp}} r_{\mathrm{comp}},
\end{split}
\end{equation}

where \(w_{\mathrm{lat}}\), \(w_{\mathrm{split}}\), \(w_{\mathrm{prog}}\),
\(w_{\mathrm{look}}\), and \(w_{\mathrm{comp}}\) are tunable weights. We adopt this multi-term reward formulation to encourage balanced execution across split and merge operations. While simpler single-term rewards can perform competitively, incorporating balanced shaping provides the most stable learning dynamics and consistent policy convergence. Below, we describe each term in detail.

\textbf{Communication Latency Reward (\(r_{\mathrm{lat}}\))}: To discourage communication schedules that increase overall delay, we penalize increases in communication makespan between consecutive decision steps. Let \(T_{\mathrm{comm}}^t\) denote the communication makespan after step \(t\):

\begin{equation}
r_{\mathrm{lat}} = -\left( T_{\mathrm{comm}}^{t+1} - T_{\mathrm{comm}}^{t} \right).
\end{equation}

Under the round-based source-channel communication model, this term is
nonzero only when a split action is scheduled from a source QPU with no
remaining active channel, causing that QPU's communication clock to
advance.

\textbf{Split Penalty (\(r_{\mathrm{split}}\))}: Since inter-QPU communication is substantially more expensive than local
computation, each split action is penalized directly:

\begin{equation}
r_{\mathrm{split}} = -\mathbb{I}_{\mathrm{split}},
\end{equation}

where \(\mathbb{I}_{\mathrm{split}}\) equals 1 when the selected action is
a split operation and 0 otherwise.

\textbf{Compilation Progress Reward (\(r_{\mathrm{prog}}\))}: To encourage continuous circuit execution, we reward the number of gates executed after each transition. Let \(\mathcal{G}_{\mathrm{fired}}^t\) denote the set of gates executed following action \(a_t\):

\begin{equation}
r_{\mathrm{prog}} = \frac{ |\mathcal{G}_{\mathrm{fired}}^t| }{ |\mathcal{G}| }.
\end{equation}

\textbf{Lookahead Reward (\(r_{\mathrm{look}}\))}: Some communication actions do not immediately execute gates but improve future execution opportunities by increasing qubit locality. We reward increases in the number of executable $CP$ gates after each
transition. Let \(\mathcal{G}_{\mathrm{exec}}^t\) denote the set of executable CP gates at step \(t\):

\begin{equation}
r_{\mathrm{look}} = |\mathcal{G}_{\mathrm{exec}}^{t+1}| - |\mathcal{G}_{\mathrm{exec}}^t|.
\end{equation}

This term encourages the policy to anticipate future execution opportunities rather than optimizing only immediate progress.

\textbf{Completion Reward (\(r_{\mathrm{comp}}\))}: A terminal reward reinforces globally successful compilation strategies, combining a completion bonus with a final communication latency penalty:

\begin{equation}
r_{\mathrm{comp}} = B_{\mathrm{complete}} \mathbb{I}_{\mathrm{complete}} - \kappa T_{\mathrm{comm}}^{\mathrm{final}},
\end{equation}

where \(B_{\mathrm{complete}}\) is the terminal completion bonus,
\(\mathbb{I}_{\mathrm{complete}}\) equals 1 when all gates in
\(\mathcal{G}\) have been executed, and \(T_{\mathrm{comm}}^{\mathrm{final}}\)
denotes the final communication makespan.

\subsection{Graph-Based State Representation}

The distributed compilation state contains multiple interacting entities, including QPUs, logical qubits, and circuit operations. These entities are coupled through hardware connectivity, qubit placement, communication-resource availability, and circuit-dependency constraints. To preserve these structured relationships, we represent each compiler state as a heterogeneous graph and process it using a graph neural network policy.

At decision step \(t\), the environment state is encoded as

\begin{equation}
\mathcal{G}^{\mathrm{obs}}_t
=
(\mathcal{V}_t,\mathcal{E}_t),
\end{equation}

where \(\mathcal{V}_t\) are typed nodes and \(\mathcal{E}_t\) are typed edges. The graph is reconstructed at every step from the current environment state.

The representation contains three node types and five edge types. Together, they capture hardware topology, qubit placement, circuit structure, communication-resource state, and execution feasibility. We summarize the key aspects of the node and edge features, with full details given in Appendix~\ref{app:features}.

\subsubsection{Node Types and Features (Table~\ref{tab:node_features}):}

The heterogeneous graph contains three categories of nodes: 

\begin{itemize}
    \item \textbf{QPU nodes:} represent hardware resources, communication state, and topology information.
    \item \textbf{Logical-qubit nodes:} represent qubit placement, communication relevance, and pending operation structure.
    \item \textbf{Gate nodes:} represent circuit operations, dependency structure, and execution readiness.
\end{itemize}

\subsubsection{Edge Types and Features (Table~\ref{tab:edge_features})}

Edges encode relationships among hardware resources, logical qubits, and circuit operations. The edge set is typed according to the relation being represented.

\subsubsection{Representation Rationale}

The heterogeneous graph preserves the relational structure of distributed compilation, allowing the policy to jointly reason about communication topology, qubit placement, resource availability, and circuit dependencies. Compared with fixed flat-vector encodings, this graph representation more naturally supports variable circuit sizes and hardware configurations while maintaining explicit relationships among QPUs, logical qubits, and gates.

\subsection{Policy Architecture}

For each edge relation \(r\), attention-based aggregation is performed as

\begin{equation}
\mathbf{m}_{u \rightarrow v}^{(r)} = \text{GAT}^{(r)}(\mathbf{h}_u, \mathbf{h}_v, \mathbf{e}_{uv}),
\end{equation}

where \(\mathbf{e}_{uv}\) denotes edge features associated with relation \(r\). Messages from all edge types are summed at each target node:

\begin{equation}
\mathbf{h}_v^{(l+1)} = \text{LayerNorm}
\left(
\mathbf{h}_v^{(l)} +
\sum_{r \in \mathcal{R}}
\sum_{u \in \mathcal{N}_r(v)}
\mathbf{m}_{u \rightarrow v}^{(r)}
\right),
\end{equation}

followed by ELU activation and dropout. The encoder outputs latent embeddings for all node types:

\[
\mathbf{h}_{\text{QPU}}, \; \mathbf{h}_{\text{qubit}}, \; \mathbf{h}_{\text{gate}}.
\]

\subsubsection{Global Context Representation}

To summarize the overall compilation state, a global context vector is constructed by pooling node embeddings across all node types. For each node category, both mean pooling and max pooling are applied:

\begin{equation}
\mathbf{g} = \psi
\left(
\text{Pool}(\mathbf{h}_{\text{QPU}})
\; \Vert \;
\text{Pool}(\mathbf{h}_{\text{qubit}})
\; \Vert \;
\text{Pool}(\mathbf{h}_{\text{gate}})
\right),
\end{equation}

where \(\Vert\) denotes concatenation and \(\psi\) is a learned linear projection. The resulting global context vector provides a compact summary of topology state, communication resources, and circuit progress.

\subsubsection{Hierarchical Action Decoder}

The action space introduced in Section~4.1 is combinatorial, since a valid communication action requires selecting a qubit, source QPU, action type, and optionally a destination QPU. To reduce complexity, action selection is factorized autoregressively into four sequential decisions:

\begin{equation}
P(a|s) =
P(q|s)
\cdot
P(p_{\text{src}}|q,s)
\cdot
P(t|q,p_{\text{src}},s)
\cdot
P(p_{\text{dst}}|q,p_{\text{src}},t,s).
\end{equation}

The decoding process proceeds as follows:

\begin{enumerate}
    \item Select a logical qubit.
    \item Select a valid source QPU currently hosting the qubit.
    \item Select an action type (Split or Merge).
    \item Select a destination QPU for split actions.
\end{enumerate}

Each decision stage is implemented using a lightweight multilayer perceptron (MLP) conditioned on node embeddings and the global context vector. At every decoding step, invalid actions are removed using feasibility masks derived from the system constraints in Section~3. This masking ensures that only valid communication actions are assigned nonzero probability. The hierarchical decomposition significantly reduces the effective branching factor and enables efficient exploration of large action spaces.

\subsection{Training Strategy}

\subsubsection{Proximal Policy Optimization (PPO)}

The policy is trained using PPO~\cite{ppo_algorithm}, a policy-gradient method that provides stable updates for sequential decision-making problems with large action spaces. PPO is well-suited for distributed compilation because it supports stochastic policies and remains robust under delayed rewards. At each training iteration, trajectories are collected by executing the current policy in the distributed compilation environment. Policy parameters are then updated using the PPO clipped objective:

\begin{equation}
L^{\mathrm{PPO}}(\theta)
=
\mathbb{E}_t
\left[
\min
\left(
\rho_t(\theta)\hat{A}_t,
\mathrm{clip}
\left(
\rho_t(\theta),
1-\epsilon,
1+\epsilon
\right)
\hat{A}_t
\right)
\right],
\end{equation}

where
\(
\rho_t(\theta)
\)
is the policy probability ratio,

\begin{equation}
\rho_t(\theta)
=
\frac{
\pi_{\theta}(a_t|s_t)
}{
\pi_{\theta_{\mathrm{old}}}(a_t|s_t)
},
\end{equation}

and \(\hat{A}_t\) is the estimated advantage. Generalized Advantage Estimation and entropy regularization are used to stabilize learning and encourage exploration.

\subsubsection{Curriculum Learning}

To improve training stability, we adopt a curriculum learning strategy that gradually increases problem complexity. Training begins with smaller circuits and progressively introduces larger circuits with deeper dependency structures and more complex communication requirements. Curriculum progression is determined by policy performance, allowing the model to first learn basic communication behaviors before addressing more challenging distributed compilation scenarios.

\section{Experiments}
\label{sec:experiment}
This section describes the experimental configuration used to evaluate the proposed reinforcement learning framework for distributed quantum compilation. The evaluation focuses on measuring whether the learned policy can reduce communication overhead relative to existing distributed compilation approaches.

\subsection{Experimental Goals}

The primary objective of the experiments is to evaluate whether the proposed RL framework can learn effective communication-aware compilation strategies for distributed quantum circuits. Rather than directly optimizing handcrafted heuristics, the framework learns scheduling behavior through interaction with the distributed execution environment.

We investigate whether the learned policy can achieve communication efficiency comparable to existing heuristic approaches while generalizing across different circuit classes and hardware configurations.

\subsection{Benchmark Circuits and Hardware Configurations}

We evaluate the framework using a collection of benchmark circuits from Qiskit \cite{qiskit}. The benchmark suite includes:

\begin{itemize}
    \item \textbf{QAOA circuits\cite{farhi2014quantum}:} generated using the Qiskit circuit library.
    \item \textbf{Quantum Fourier Transform (QFT)\cite{weinstein2001implementation}:} generated using Qiskit reference implementations.
    \item \textbf{Quantum Volume circuits\cite{cross2019validating}:} generated using the Qiskit benchmarking suite.
    \item \textbf{Random circuits\cite{burt2025multilevel}:} generated with fixed depth using the \texttt{cp\_fraction} circuit~\cite{burt2025multilevel}.
\end{itemize}

All experiments are conducted using two-QPU distributed architectures under two hardware configurations. All circuits are compiled into controlled-phase gates $CP(\theta)$ and single-qubit parameterized rotations $U(\phi,\theta,\lambda)$ using the Qiskit transpiler~\cite{qiskit}.

\begin{table}[!b]
\centering
\caption{Distributed hardware configurations.}
\label{tab:hardware_config}
\small
\renewcommand{\arraystretch}{1.15}
\setlength{\tabcolsep}{4pt}
\begin{tabular}{@{}lccc@{}}
\hline
\textbf{Experiments} & \textbf{QPUs} & \shortstack{\textbf{Qubits / QPU}} & \shortstack{\textbf{Channel Capacity}} \\
\hline
Section~\ref{sec:depth_exp}& 2 & 4 & 2 \\
Section~\ref{sec:benchmark_exp} & 2 & 8 & 4 \\
\hline
\end{tabular}
\end{table}

\subsection{Baseline and Evaluation Metric}

We compare the proposed framework against two software libraries Pytket-DQC~\cite{andres-martinezCQCLPytketdqc2024} and DISQCO~\cite{burt2025disqco}. Pytket-DQC is based on Refs.~\cite{andres2019automated, wu2023entanglement, andres2024distributing}, while DISQCO is based on Refs.~\cite{burt2025multilevel,burt2025entanglement}. Both libraries used graph-based heuristics for minimizing the number of EPR pairs consumed when distributing quantum circuits under network connectivity constraints. Since these methods optimise only for EPR count, performance is evaluated using the total number of split operations introduced during compilation, which directly corresponds to the number of EPR generation events. Importantly, while Pytket-DQC and DISQCO assume unconstrained link availability, our framework enforces explicit QPU channel capacities to account for physical bandwidth contention. This introduces a more stringent execution model without altering the underlying EPR count logic. From DISQCO, we use the default multilevel partitioner, \texttt{MLFM-R}~\cite{burt2025multilevel}. From Pyktet-DQC, we use the heterogeneous hypergraph partitioner, \texttt{Partition}~\cite{andres2024distributing}.



\begin{table}[h]
\centering
\caption{Reward hyperparameters used in all experiments.}
\label{tab:reward_hyperparameters}
\small
\renewcommand{\arraystretch}{1.15}
\setlength{\tabcolsep}{5pt}
\begin{tabular}{lc}
\hline
\textbf{Hyperparameter} & \textbf{Value} \\
\hline
Latency reward weight \(w_{\mathrm{lat}}\) & 1.0 \\
Split penalty weight \(w_{\mathrm{split}}\) & 0.3 \\
Progress reward weight \(w_{\mathrm{prog}}\) & 20.0 \\
Lookahead reward weight \(w_{\mathrm{look}}\) & 0.5 \\
Completion reward weight \(w_{\mathrm{comp}}\) & 1.0 \\
Completion bonus \(B_{\mathrm{complete}}\) & 10.0 \\
Terminal latency penalty \(\kappa\) & 0.05 \\
\hline
\end{tabular}
\end{table}

The reward weights shown in Table~\ref{tab:reward_hyperparameters} are fixed across all experiments. These values were selected empirically to balance short-term circuit progress with long-term communication efficiency.

\section{Results and Analysis}
\label{sec:wireless-model}
This section evaluates the proposed reinforcement learning framework across multiple circuit classes and hardware configurations. We compare the learned policy against DISQCO and Pytket-DQC using communication cost measured by split operations, which correspond directly to EPR generation events.

The analysis focuses on two aspects: (1) overall competitiveness relative to heuristic compilation and (2) behavior across circuit complexity and benchmark families. Through these experiments, we aim to understand when reinforcement learning is effective for distributed quantum compilation and identify scenarios where learned policies remain challenging.

\subsection{Effect of Circuit Structure on Communication Cost}
\label{sec:circuit_structure_cost}

\label{sec:depth_exp}

\begin{figure*}
    \centering
    \begin{tikzpicture}
	\begin{axis}[
		width=0.48\textwidth,
		height=0.38\textwidth,
		xlabel={circuit depth},
		ylabel={communication cost (\# EPR)},
		title={Random},
		legend pos=north west,
		grid=major,
		xmin=20, xmax=200,
	]
		\addplot[color=blue, mark=*, mark size=1.2pt] table[x=depth,y=DISQCO,col sep=comma] {plots/random_circs.csv};
		\addlegendentry{DISQCO}
    
        \addplot[color=orange, mark=*, mark size=1.2pt] table[x=depth,y=pytket,col sep=comma] {plots/random_circs.csv};
		\addlegendentry{Pytket}

		\addplot[color=red, mark=square*, mark size=1.2pt] table[x=depth,y=RL,col sep=comma] {plots/random_circs.csv};
		\addlegendentry{RL}
	\end{axis}
\end{tikzpicture}
    \begin{tikzpicture}
	\begin{axis}[
		width=0.48\textwidth,
		height=0.38\textwidth,
		xlabel={circuit depth},
		ylabel={communication cost (\# EPR)},
		title={Quantum Volume},
		legend pos=north west,
		grid=major,
		xmin=20, xmax=200,
	]
		\addplot[color=blue, mark=*, mark size=1.2pt] table[x=depth,y=DISQCO,col sep=comma] {plots/qv_circs.csv};
		\addlegendentry{DISQCO}

        \addplot[color=orange, mark=*, mark size=1.2pt] table[x=depth,y=pytket,col sep=comma] {plots/qv_circs.csv};
		\addlegendentry{Pytket}

		\addplot[color=red, mark=square*, mark size=1.2pt] table[x=depth,y=RL,col sep=comma] {plots/qv_circs.csv};
		\addlegendentry{RL}
	\end{axis}
\end{tikzpicture}
    \caption{Communication cost versus circuit depth for random and quantum volume benchmarks.}
    \label{fig:depth_cost}
\end{figure*}
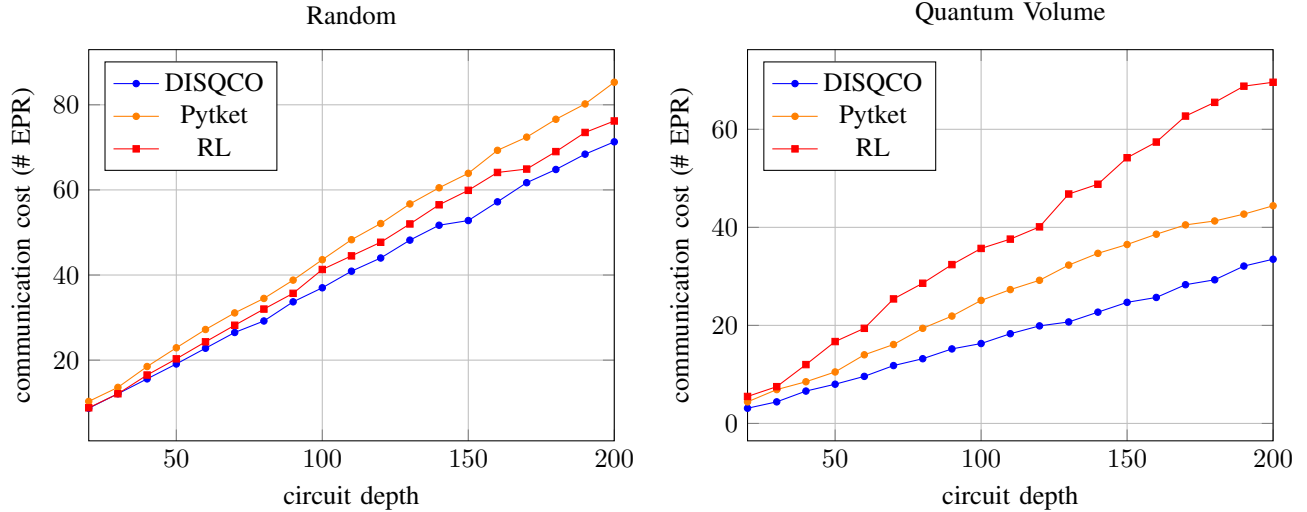

We examine how communication cost changes with circuit depth under the 2-QPU, 4-qubit/QPU configuration. Figure~\ref{fig:depth_cost} compares the number of split operations required by the proposed RL framework, DISQCO and Pytket-DQC, on random and quantum volume benchmarks. For random circuits, we sample circuits with a $0.5-0.8$ fraction of $CP$ gates, sweeping the depth. For quantum volume, we generate circuits with multiple layers, then decompose these into $CP$ and $U$ gates. We then sweep the depth by truncating the transpiled depth.

For random circuits, the learned RL policy achieves a lower communication cost than Pytket-DQC and remains comparable to DISQCO across the evaluated depth range. This indicates that the policy can learn effective communication decisions when two-qubit interactions are relatively sparse and distributed across the circuit. In such cases, a split or merge action often has a clearer local effect on near-term gate executability, making the reward signal easier to associate with useful communication decisions.

For quantum volume circuits, however, the RL policy requires approximately twice as many split operations as DISQCO. This behavior can be explained by the denser interaction structure of quantum volume circuits. Each quantum volume layer induces many two-qubit interactions through qubit permutations, often involving more logical qubits than can fit within a single QPU. Consequently, the compiler must coordinate communication decisions across a larger interacting region. Because many qubits simultaneously appear important for future execution, the action-value signal for selecting one specific split or merge becomes less distinguishable. The policy may therefore choose locally reasonable but globally redundant communication actions, leading to higher split counts than DISQCO.

\subsection{Benchmark-Level Comparison and Ablation}

\label{sec:benchmark_exp}

To evaluate performance across circuit families, Table~\ref{tab:weak_baseline_ablation} reports the average communication cost for QAOA, QFT, random, and quantum volume benchmarks. These circuits are generated using $12$ qubits targeted for a 2-QPU topology with $8$ qubits on each QPU. Communication cost is measured by the total number of split operations, which represents the EPR count. Lower values indicate fewer inter-QPU communication events.

In addition to the two heuristics from the literature, we compare our complete RL-framework (\textit{Full RL)} against two RL reward ablations. \textit{Progress-only RL} removes the lookahead reward component and relies purely on immediate gate execution feedback, while \textit{No-lookahead RL} evaluates a policy trained without future-horizon state lookahead.

\begin{table}[t]
\centering
\caption{Average number of split operations across benchmark families. Lower is better.}
\label{tab:weak_baseline_ablation}
\begin{tabular}{lcccc}
\toprule
Method & QAOA & QFT & Random & QV\\
\midrule
Progress-only RL & 4.0 & 4.0 & 42.0 & 29.0 \\
No-lookahead RL & 4.0 & 4.0 & 40.0 & 24.7 \\
\textbf{Full RL} & \textbf{4.0} & \textbf{4.0} & \textbf{39.0} & \textbf{28.7}  \\
Pytket-DQC & 3.0 & 5.7 & 39.5 & 17.5 \\
DISQCO & 3.0 & 4.0 & 35.7 & 11.9 \\
\bottomrule
\end{tabular}
\end{table}

For QAOA and QFT, the absolute communication cost is small across all methods. These benchmarks are relatively shallow in the evaluated setting, and the optimal communication strategy is often simple~\cite{burt2025multilevel}. Still, the RL model is able to discover these strategies is a good indicator of its performance under structured circuits.

Quantum Volume shows the largest gap, with \textit{Full RL} requiring \(28.7\) splits compared with \(11.9\) for DISQCO. As discussed in Section~\ref{sec:circuit_structure_cost}, this reflects the difficulty of learning globally coordinated communication decisions in dense interaction regions.

Comparing reward variants, \textit{Full RL} achieves communication costs comparable to \textit{Progress-only} and \textit{No-lookahead RL}. On random circuits, \textit{Full RL} offers a modest reduction in average split count compared to the alternative reward formulations ($39.0$ versus $42.0$ for \textit{Progress-only} and $40.0$ for \textit{No-lookahead}). However, the overall improvements across these RL variants remain incremental, with \textit{No-lookahead} actually outperforming \textit{Full RL} on Quantum Volume ($24.7$ vs. $28.7$). This suggests that while lookahead rewards can slightly assist communication decisions with delayed benefits in unstructured settings, the optimal reward structure depends heavily on the underlying circuit architecture.

Overall, \textit{Full RL} benefits from lookahead guidance on random circuits and remains competitive on structured benchmarks, though further refinements are needed for dense interacting circuits.

\section{Conclusion}
\label{sec:conclusion}
This work formulates distributed quantum circuit compilation as a communication-aware sequential decision-making problem. It presents a reinforcement learning framework for scheduling inter-QPU communication in distributed quantum architectures. By modeling compilation as a Markov Decision Process and encoding compiler states as heterogeneous graphs, the proposed approach captures the coupled structure of circuit dependencies, logical-qubit placement, and hardware constraints. Experimental results show that the learned policy achieves communication costs comparable to state-of-the-art heuristics on structured benchmark families like QFT and QAOA. Evaluating alternative reward design variants indicates that lookahead reward shaping provides modest performance gains on unstructured random circuits by encouraging decisions with delayed benefits. However, highly interacting circuits remain challenging. These findings demonstrate that our reinforcement learning framework offers a flexible, architecture-aware alternative to handcrafted compilation heuristics, though scalability and performance on dense workloads remain open challenges. Future work will focus on training on specific classes of circuits to identify common communication patterns, as well as improving policy guidance through global-placement heuristics, hybrid RL formulations, and richer communication-action representations.

\appendix
\section{Graph Features}
\label{app:features}
Here we specify exactly the features used in the graph representation.

\begin{table}[h]
\centering
\caption{Edge feature summary.}
\label{tab:edge_features}
\scriptsize
\setlength{\tabcolsep}{3pt}
\renewcommand{\arraystretch}{1.08}
\begin{tabular}{p{1.45cm}p{2.05cm}p{4.15cm}}
\hline
\textbf{Type} & \textbf{Feature} & \textbf{Description} \\
\hline
\multirow{1}{*}{QPU--QPU}
& Link indicator & Indicates that two QPUs are connected by a valid communication link. \\
& Topology distance & Optional normalized graph distance or adjacency feature used for topology awareness. \\
\hline
\multirow{1}{*}{Qubit--QPU}
& Presence & Indicates whether a logical qubit is currently available at a QPU. \\
& CP ready ratio & Fraction of visible CP partners that can be co-located at this QPU. \\
& Merge opportunity & Indicates whether removing this placement relation is currently feasible. \\
\hline
Gate--Qubit
& Operand type & Distinguishes single-qubit participation from two-qubit CP participation. \\
\hline
Gate--Gate
& Depth distance & Relative dependency-depth distance between connected gates. \\
\hline
Qubit--Qubit
& Co-location & Indicates whether two logical qubits currently share at least one QPU. \\
\hline
\end{tabular}
\end{table}

\begin{table}[h]
\centering
\caption{Node feature summary.}
\label{tab:node_features}
\scriptsize
\setlength{\tabcolsep}{3pt}
\renewcommand{\arraystretch}{1.08}
\begin{tabular}{p{1.25cm}p{2.15cm}p{4.25cm}}
\hline
\textbf{Type} & \textbf{Feature} & \textbf{Description} \\
\hline
\multirow{1}{*}{QPU}
& Free capacity ratio & Available qubit slots normalized by total capacity. \\
& Normalized clock & Communication clock normalized relative to the current system state. \\
& Max source channels & Maximum number of split actions that can be scheduled from the QPU within one communication round. \\
& Active source-channel ratio & Remaining source-channel budget normalized by maximum channel capacity. \\
& Degree & Normalized topology connectivity. \\
& WL topology ID & Weisfeiler--Lehman encoding of topology symmetry. \\
& Clock skew & Relative clock offset compared to the average QPU clock. \\
& Unused channel budget & Remaining communication opportunities in the current round. \\
\hline
\multirow{1}{*}{Logical qubit}
& \# locations & Number of QPUs where the qubit is currently available. \\
& CP blocked ratio & Fraction of visible CP gates blocked by placement mismatch. \\
& Criticality & Relative importance based on dependency depth or critical-path participation. \\
& CP count share & Relative participation in visible CP operations. \\
& Ready 1Q & Indicates whether a dependency-satisfied single-qubit gate is executable. \\
& Has 1Q & Indicates whether pending single-qubit operations remain. \\
& Has merge & Indicates whether a merge action is currently valid. \\
& Pending split & Indicates a recently created split not yet utilized by execution. \\
& Packet root weight & Importance within grouped CP communication packets. \\
\hline
\multirow{1}{*}{Gate}
& Is CP & Binary indicator for a CP gate. \\
& Is 1Q & Binary indicator for a single-qubit gate. \\
& Normalized depth & Relative position within the compilation window. \\
& Max depth flag & Indicates critical-depth membership. \\
& Predecessor count & Normalized dependency fan-in. \\
& Successor count & Normalized dependency fan-out. \\
& Blocked successors & Fraction of downstream gates blocked by this gate. \\
& Frontier flag & Indicates whether the gate lies on the current executable frontier. \\
& Placement satisfied & Indicates whether placement constraints are currently satisfied. \\
& CP placement blocked & Indicates whether a CP gate is blocked by qubit placement. \\
& Packet size & Size of the associated CP communication packet. \\
& Packet root & Indicates leader gate within a packet group. \\
\hline
\end{tabular}
\end{table}

\section*{Acknowledgment}
This work was supported by the Engineering and Physical Sciences Research Council (EPSRC) under grant number EP/W032643/1.

\bibliographystyle{ieeetr}
\bibliography{references}

@article{wu2023entanglement,
  title = {Entanglement-Efficient Bipartite-Distributed Quantum Computing},
  author = {Wu, Jun-Yi and Matsui, Kosuke and Forrer, Tim and Soeda, Akihito and {Andr{\'e}s-Mart{\'i}nez}, Pablo and Mills, Daniel and Henaut, Luciana and Murao, Mio},
  year = 2023,
  month = dec,
  journal = {Quantum},
  volume = {7},
  eprint = {2212.12688},
  primaryclass = {quant-ph},
  pages = {1196},
  issn = {2521-327X},
  doi = {10.22331/q-2023-12-05-1196},
  urldate = {2024-06-04},
  archiveprefix = {arXiv},
}

@misc{yimsiriwattana2004generalized,
  title = {Generalized {{GHZ States}} and {{Distributed Quantum Computing}}},
  author = {Yimsiriwattana, Anocha and Lomonaco Jr, Samuel J.},
  year = 2004,
  month = mar,
  number = {arXiv:quant-ph/0402148},
  eprint = {quant-ph/0402148},
  publisher = {arXiv},
  urldate = {2023-11-28},
  archiveprefix = {arXiv},
}

@article{andres2024distributing,
  title = {Distributing Circuits over Heterogeneous, Modular Quantum Computing Network Architectures},
  author = {{Andres-Martinez}, Pablo and Forrer, Tim and Mills, Daniel and Wu, Jun-Yi and Henaut, Luciana and Yamamoto, Kentaro and Murao, Mio and Duncan, Ross},
  year = 2024,
  month = aug,
  journal = {Quantum Science and Technology},
  volume = {9},
  number = {4},
  pages = {045021},
  publisher = {IOP Publishing},
  issn = {2058-9565},
  doi = {10.1088/2058-9565/ad6734},
  urldate = {2024-08-22},
  langid = {english},
}

@article{caleffi2024distributed,
  title={Distributed quantum computing: a survey},
  author={Caleffi, Marcello and Amoretti, Michele and Ferrari, Davide and Illiano, Jessica and Manzalini, Antonio and Cacciapuoti, Angela Sara},
  journal={Computer Networks},
  volume={254},
  pages={110672},
  year={2024},
  publisher={Elsevier}
}

@article{zulehner2018efficient,
  title={An efficient methodology for mapping quantum circuits to the IBM QX architectures},
  author={Zulehner, Alwin and Paler, Alexandru and Wille, Robert},
  journal={IEEE Transactions on Computer-Aided Design of Integrated Circuits and Systems},
  volume={38},
  number={7},
  pages={1226--1236},
  year={2018},
  publisher={IEEE}
}

@inproceedings{mao2023qubit,
  title={Qubit allocation for distributed quantum computing},
  author={Mao, Yingling and Liu, Yu and Yang, Yuanyuan},
  booktitle={IEEE INFOCOM 2023-IEEE Conference on Computer Communications},
  pages={1--10},
  year={2023},
  organization={IEEE}
}

@inproceedings{wu2023qucomm,
  title={Qucomm: Optimizing collective communication for distributed quantum computing},
  author={Wu, Anbang and Ding, Yufei and Li, Ang},
  booktitle={Proceedings of the 56th Annual IEEE/ACM International Symposium on Microarchitecture},
  pages={479--493},
  year={2023}
}

@inproceedings{wu2022autocomm,
  title={Autocomm: A framework for enabling efficient communication in distributed quantum programs},
  author={Wu, Anbang and Zhang, Hezi and Li, Gushu and Shabani, Alireza and Xie, Yuan and Ding, Yufei},
  booktitle={2022 55th IEEE/ACM International Symposium on Microarchitecture (MICRO)},
  pages={1027--1041},
  year={2022},
  organization={IEEE}
}

@article{sundaram2022distribution,
  title={Distribution of quantum circuits over general quantum networks},
  author={Sundaram, Ranjani G and Gupta, Himanshu and Ramakrishnan, CR},
  journal={arXiv preprint arXiv:2206.06437},
  year={2022}
}

@article{ferrari2023modular,
  title={A modular quantum compilation framework for distributed quantum computing},
  author={Ferrari, Davide and Carretta, Stefano and Amoretti, Michele},
  journal={IEEE Transactions on Quantum Engineering},
  volume={4},
  pages={1--13},
  year={2023},
  publisher={IEEE}
}

@article{cuomo2023optimized,
  title={Optimized compiler for distributed quantum computing},
  author={Cuomo, Daniele and Caleffi, Marcello and Krsulich, Kevin and Tramonto, Filippo and Agliardi, Gabriele and Prati, Enrico and Cacciapuoti, Angela Sara},
  journal={ACM Transactions on Quantum Computing},
  volume={4},
  number={2},
  pages={1--29},
  year={2023},
  publisher={ACM New York, NY}
}

@article{andres2019automated,
  title={Automated distribution of quantum circuits via hypergraph partitioning},
  author={Andres-Martinez, Pablo and Heunen, Chris},
  journal={Physical Review A},
  volume={100},
  number={3},
  pages={032308},
  year={2019},
  publisher={APS}
}

@article{childs2019circuit,
  title={Circuit transformations for quantum architectures},
  author={Childs, Andrew M and Schoute, Eddie and Unsal, Cem M},
  journal={arXiv preprint arXiv:1902.09102},
  year={2019}
}

@article{siraichi2019qubit,
  title={Qubit allocation as a combination of subgraph isomorphism and token swapping},
  author={Siraichi, Marcos Yukio and Santos, Vin{\'\i}cius Fernandes dos and Collange, Caroline and Pereira, Fernando Magno Quint{\~a}o},
  journal={Proceedings of the ACM on Programming Languages},
  volume={3},
  number={OOPSLA},
  pages={1--29},
  year={2019},
  publisher={ACM New York, NY, USA}
}

@inproceedings{shende2005synthesis,
  title={Synthesis of quantum logic circuits},
  author={Shende, Vivek V and Bullock, Stephen S and Markov, Igor L},
  booktitle={Proceedings of the 2005 Asia and South Pacific Design Automation Conference},
  pages={272--275},
  year={2005}
}

@article{cacciapuoti2019quantum,
  title={Quantum internet: Networking challenges in distributed quantum computing},
  author={Cacciapuoti, Angela Sara and Caleffi, Marcello and Tafuri, Francesco and Cataliotti, Francesco Saverio and Gherardini, Stefano and Bianchi, Giuseppe},
  journal={IEEE Network},
  volume={34},
  number={1},
  pages={137--143},
  year={2019},
  publisher={IEEE}
}

@article{eisert2000optimal,
  title={Optimal local implementation of nonlocal quantum gates},
  author={Eisert, Jens and Jacobs, Kurt and Papadopoulos, Polykarpos and Plenio, Martin B},
  journal={Physical Review A},
  volume={62},
  number={5},
  pages={052317},
  year={2000},
  publisher={APS}
}

@article{zukowski1993event,
  title={‘‘Event-ready-detectors’’Bell experiment via entanglement swapping},
  author={{\.Z}ukowski, Marek and Zeilinger, Anton and Horne, Michael A and Ekert, Aarthur K},
  journal={Physical review letters},
  volume={71},
  number={26},
  pages={4287},
  year={1993},
  publisher={APS}
}

@article{bennett1993teleporting,
  title={Teleporting an unknown quantum state via dual classical and Einstein-Podolsky-Rosen channels},
  author={Bennett, Charles H and Brassard, Gilles and Cr{\'e}peau, Claude and Jozsa, Richard and Peres, Asher and Wootters, William K},
  journal={Physical review letters},
  volume={70},
  number={13},
  pages={1895},
  year={1993},
  publisher={APS}
}

@article{cross2019validating,
  title={Validating quantum computers using randomized model circuits},
  author={Cross, Andrew W and Bishop, Lev S and Sheldon, Sarah and Nation, Paul D and Gambetta, Jay M},
  journal={Physical Review A},
  volume={100},
  number={3},
  pages={032328},
  year={2019},
  publisher={APS}
}

@article{weinstein2001implementation,
  title={Implementation of the quantum Fourier transform},
  author={Weinstein, Yaakov S and Pravia, MA and Fortunato, EM and Lloyd, Seth and Cory, David G},
  journal={Physical review letters},
  volume={86},
  number={9},
  pages={1889},
  year={2001},
  publisher={APS}
}

@article{farhi2014quantum,
  title={A quantum approximate optimization algorithm},
  author={Farhi, Edward and Goldstone, Jeffrey and Gutmann, Sam},
  journal={arXiv preprint arXiv:1411.4028},
  year={2014}
}

@article{van2016path,
  title={The path to scalable distributed quantum computing},
  author={Van Meter, Rodney and Devitt, Simon J},
  journal={Computer},
  volume={49},
  number={9},
  pages={31--42},
  year={2016},
  publisher={IEEE}
}

@article{cirac1999distributed,
  title={Distributed quantum computation over noisy channels},
  author={Cirac, J Ignacio and Ekert, AK and Huelga, Susana F and Macchiavello, Chiara},
  journal={Physical Review A},
  volume={59},
  number={6},
  pages={4249},
  year={1999},
  publisher={APS}
}

@inproceedings{burt2024generalised,
  title={Generalised circuit partitioning for distributed quantum computing},
  author={Burt, Felix and Chen, Kuan-Cheng and Leung, Kin K},
  booktitle={2024 IEEE International Conference on Quantum Computing and Engineering (QCE)},
  volume={2},
  pages={173--178},
  year={2024},
  organization={IEEE}
}

@article{burt2025multilevel,
  title={A multilevel framework for partitioning quantum circuits},
  author={Burt, Felix and Chen, Kuan-Cheng and Leung, Kin K},
  journal={Quantum},
  volume={10},
  pages={1984},
  year={2026},
  publisher={Verein zur F{\"o}rderung des Open Access Publizierens in den Quantenwissenschaften}
}

@article{chen2026adaptive,
  title={Adaptive Resource Orchestration for Distributed Quantum Computing Systems},
  author={Chen, Kuan-Cheng and Burt, Felix and Panigrahy, Nitish K and Leung, Kin K},
  journal={IEEE Internet Computing},
  year={2026},
  publisher={IEEE}
}

@article{wootters1982single,
  title={A single quantum cannot be cloned},
  author={Wootters, William K and Zurek, Wojciech H},
  journal={Nature},
  volume={299},
  number={5886},
  pages={802--803},
  year={1982},
  publisher={Nature Publishing Group UK London}
}

@inproceedings{burt2025entanglement,
  title={Entanglement-efficient distribution of quantum circuits over large-scale quantum networks},
  author={Burt, Felix and Chen, Kuan-Cheng and Leung, Kin K},
  booktitle={2025 IEEE International Conference on Quantum Computing and Engineering (QCE)},
  volume={1},
  pages={1111--1122},
  year={2025},
  organization={IEEE}
}

@misc{andres-martinezCQCLPytketdqc2024,
  title = {{{CQCL}}/Pytket-Dqc},
  author = {{Andres-Martinez}, Pablo and Mills, Daniel and Forrer, Tim and Henaut, Luciana},
  year = {2024},
  publisher = {GitHub},
  journal = {GitHub repository},
  howpublished = {\url{https://github.com/Quantinuum/pytket-dqc}}
}

@misc{burt2025disqco,
  author = {Burt, Felix},
  title = {DISQCO: Distributed quantum circuit optimisation},
  year = {2025},
  publisher = {GitHub},
  journal = {GitHub repository},
  howpublished = {\url{https://github.com/felix-burt/DISQCO}},
}

@misc{ppo_algorithm,
      title={Proximal Policy Optimization Algorithms}, 
      author={John Schulman and Filip Wolski and Prafulla Dhariwal and Alec Radford and Oleg Klimov},
      year={2017},
      eprint={1707.06347},
      archivePrefix={arXiv},
      primaryClass={cs.LG},
      url={https://arxiv.org/abs/1707.06347}, 
}

@misc{qiskit,
      title={Quantum computing with Qiskit}, 
      author={Ali Javadi-Abhari and Matthew Treinish and Kevin Krsulich and Christopher J. Wood and Jake Lishman and Julien Gacon and Simon Martiel and Paul D. Nation and Lev S. Bishop and Andrew W. Cross and Blake R. Johnson and Jay M. Gambetta},
      year={2024},
      eprint={2405.08810},
      archivePrefix={arXiv},
      primaryClass={quant-ph},
      url={https://arxiv.org/abs/2405.08810}, 
}

@Inbook{rl_mdp,
author="van Otterlo, Martijn
and Wiering, Marco",
editor="Wiering, Marco
and van Otterlo, Martijn",
title="Reinforcement Learning and Markov Decision Processes",
bookTitle="Reinforcement Learning: State-of-the-Art",
year="2012",
publisher="Springer Berlin Heidelberg",
address="Berlin, Heidelberg",
pages="3--42",
isbn="978-3-642-27645-3",
doi="10.1007/978-3-642-27645-3_1",
url="https://doi.org/10.1007/978-3-642-27645-3_1"
}

@inproceedings{promponas2025compiler,
  author    = {Panagiotis Promponas and Akrit Mudvari and Luca Della Chiesa
               and Paul Polakos and Louis Samuel and Leandros Tassiulas},
  title     = {Compiler for Distributed Quantum Computing:
               A Reinforcement Learning Approach},
  booktitle = {ICC 2025--IEEE International Conference on Communications},
  pages     = {4615--4621},
  year      = {2025},
  publisher = {IEEE},
  doi       = {10.1109/ICC52391.2025.11161115}
}

@article{escofet2025revisiting,
  author  = {Pau Escofet and Anabel Ovide and Medina Bandic
             and Luise Prielinger and Hans van Someren and Sebastian Feld
             and Eduard Alarc{\'o}n and Sergi Abadal
             and Carmen G. Almud{\'e}ver},
  title   = {Revisiting the Mapping of Quantum Circuits:
             Entering the Multi-Core Era},
  journal = {ACM Transactions on Quantum Computing},
  volume  = {6},
  number  = {1},
  articleno = {4},
  pages   = {1--26},
  year    = {2025},
  doi     = {10.1145/3655029}
}

@inproceedings{russo2025telesabre,
  author    = {Enrico Russo and Elio Vinciguerra and Maurizio Palesi
               and Davide Patti and Giuseppe Ascia and Vincenzo Catania},
  title     = {{TeleSABRE}: Heuristic Layout Synthesis in Multi-Core
               Quantum Systems with Teleport Interconnect},
  booktitle = {2025 IEEE International Conference on Quantum Computing
               and Engineering (QCE)},
  year      = {2025},
  publisher = {IEEE},
  doi       = {10.1109/QCE65121.2025.00086}
}

@article{bandic2025profiling,
  author  = {Medina Bandic and Pablo le Henaff and Anabel Ovide
             and Pau Escofet and Sahar Ben Rached and Santiago Rodrigo
             and Hans van Someren and Sergi Abadal and Eduard Alarc{\'o}n
             and Carmen G. Almud{\'e}ver and Sebastian Feld},
  title   = {Profiling Quantum Circuits for Their Efficient Execution
             on Single- and Multi-Core Architectures},
  journal = {Quantum Science and Technology},
  volume  = {10},
  number  = {1},
  pages   = {015060},
  year    = {2025},
  doi     = {10.1088/2058-9565/ada180}
}

@article{benrached2025traffic,
  author  = {Sahar Ben Rached and Isaac Lopez Agudo and Santiago Rodrigo
             and Medina Bandic and Artur Garcia-Saez and Sebastian Feld
             and Hans van Someren and Eduard Alarc{\'o}n
             and Carmen G. Almud{\'e}ver and Sergi Abadal},
  title   = {Characterizing the Inter-Core Qubit Traffic in Large-Scale
             Quantum Modular Architectures},
  journal = {IEEE Access},
  volume  = {13},
  pages   = {113236--113257},
  year    = {2025},
  doi     = {10.1109/ACCESS.2025.3583218}
}

@article{pouryousef2025network,
  author  = {Shahrooz Pouryousef and Reza Nejabati and Don Towsley
             and Ramana Kompella and Eneet Kaur},
  title   = {Network-Aware Scheduling for Remote Gate Execution
             in Quantum Data Centers},
  journal = {arXiv preprint arXiv:2504.20176},
  year    = {2025}
}

@article{burt2026multilevel,
  author  = {Felix Burt and Kuan-Cheng Chen and Kin K. Leung},
  title   = {A Multilevel Framework for Partitioning Quantum Circuits},
  journal = {Quantum},
  volume  = {10},
  pages   = {1984},
  year    = {2026},
  doi     = {10.22331/q-2026-01-22-1984}
}

@INPROCEEDINGS{qsyn,
  author={Lau, Mu-Te and Cheng, Chin-Yi and Lu, Cheng-Hua and Chuang, Chia-Hsu and Kuo, Yi-Hsiang and Yang, Hsiang-Chun and Kuo, Chien-Tung and Chen, Hsin-Yu and Tung, Chen-Ying and Tsai, Cheng-En and Chen, Guan-Hao and Lin, Leng-Kai and Wang, Ching-Huan and Wang, Tzu-Hsu and Huang, Chung-Yang Ric},
  booktitle={2024 IEEE International Conference on Quantum Computing and Engineering (QCE)}, 
  title={Qsyn: A Developer-Friendly Quantum Circuit Synthesis Framework for NISQ Era and Beyond}, 
  year={2024},
  volume={02},
  number={},
  pages={535-536},
  doi={10.1109/QCE60285.2024.10392}}

\end{document}